\documentclass{article}
\usepackage{spconf,amsmath,epsfig}
\usepackage[colorlinks,linkcolor=blue, anchorcolor=blue, citecolor=blue]{hyperref}

\title{Wake effects of offshore wind farm clusters revealed by SAR and WRF}
\name{Rui Li, Jincheng Zhang, Xiaowei Zhao *}
\address{Intelligent Control \& Smart Energy (ICSE) Research Group, \\School of Engineering, University of Warwick, Coventry, CV4 7AL, UK\\
*Corresponding author: Xiaowei.Zhao@warwick.ac.uk}
\begin{document}
%
\maketitle
\begin{abstract}
Wake effects, i.e. the reduced momentum and increased turbulence caused by the upstream wind farm, have a significant adverse impact on downstream wind farms. However, due to the lack of ground truth for flow scenarios without wind farms in place (as the wind farm has already been constructed on site), it is extremely difficult to quantify the real impact caused by the presence of upstream wind farms for the downstream area. This paper seeks to develop a potential solution by taking advantage of both SAR and WRF. Specifically, the real-world wind speed with wind farms is retrieved from the SAR images using the C-band model, while the scenario without wind farms is simulated by the WRF model. By combining these two technologies, the potential impact of long-distance wind farm wakes is revealed and analysed.
\end{abstract}
\section{Introduction}
\label{sec:intro}
As a major source of sustainable energy, offshore wind has experienced substantial growth in recent decades \cite{GWEC}. When planning offshore wind farms, several factors should be taken into consideration, such as potential wind resources, overall construction cost, operation and maintenance expenses, and proximity to transmission lines and roads \cite{lundquist2019costs}. However, sites with all those favourable advantages are often limited, leading to the clustering of wind farms in close sea areas. For example, several large-scale wind farms of different countries are already operating in the North Sea, with more farms planned for the future. With more and more farm projects developed close to each other, the wake interaction between upstream and downstream farms become more and more important. The wake effects should thus be taken into account more carefully, so that the wakes' impact can be mitigated for the optimal design and operations of wind farms.

Currently, to investigate the wake effects caused by wind farms, two mainstream technologies have been used, i.e. mesoscale simulations \cite{cuevas2022accuracy, ali2023assessment} and in-situ measurements \cite{schneemann2021offshore}. However, mesoscale models cannot fully capture the multi-scale physical characteristics of wake effects. This is particularly true when considering real-world operational scenarios. When it comes to in-situ measurements, as the wind farms have already been installed in place, it is impossible to obtain the ground truth for the `farm-less' scenario. The quantification of wake effects via comparison with wind flows in the `farm-less' scenario is thus not feasible. 

In this paper, we combine the advantages of satellite observations and mesoscale simulations to visually demonstrate the wake effects of large offshore wind farm clusters. Although a series of studies have been conducted to combine the advantages of both SAR and WRF \cite{hasager2015comparing, larsen2021case}, their primary focus lies in refining WRF simulations through wind farm parameterization based on SAR-observed results. Our paper, in contrast, focuses on visualizing the wind flow characteristics between scenarios with and without wind farms. Specifically, the wind speed with wind farms is retrieved from the Synthetic Aperture Radar (SAR) images using a C-band model, i.e. CMOD5.N \cite{hersbach2010comparison}. In parallel, for the scenarios without wind farms, wind speed is simulated with the Weather Research and Forecasting (WRF) model \cite{skamarock2019description}. By comparing wind speeds in scenarios with and without wind farms, the impact of wake effects can be effectively illustrated. It is noteworthy that other closely-related works on investigating the potential impact of operational wind farms include approaches based on SAR \cite{owda2022wind} (i.e. before and after commissioning of offshore wind farms) and approaches based on WRF \cite{mayol2021farm} (i.e. with and without wind farm parameterization).

\section{Methodology}

In this paper, four typical wind farms located in distinct sea areas are selected to conduct our experiments: 1) Hornsea Project One, UK; 2) Butendiek, DE; 3) Hohe See, Albatros and Global Tech I, DE; 4) Datang Jiangsu Binhai, CN. Meanwhile, the corresponding WRF simulations are carried out according to coordinates and acquisition times. We mention that the approach in the paper is generic to other wind farm sites.

\label{sec:metho}
\subsection{Wind speed retrieval from SAR}
\label{sec:sar}
In this study, we collect the sentinel-1 images from the Copernicus Data Space Ecosystem, specifying the product type as GRD and the sensor mode set to IW. Subsequently, a comprehensive preprocessing procedure is executed using the SNAP toolbox. This includes thermal noise removal, orbit correction, radiometric calibration, speckle noise filtering, bright object removal, and multi-looking, resulting in processed images with a resolution of 500 meters. Finally, the CMOD5.N \cite{hersbach2010comparison} is used to retrieve the wind speed from the processed sentinel images. The azimuth and incidence angles are directly extracted from the Sentinel-1 images, while the wind direction is obtained from ERA5 hourly data \cite{hersbach2018era5}. More details on wind speed retrieval can be found in our parallel work \cite{li2023long}.

\subsection{Wind speed simulated by WRF}
\label{sec:wrf}
\begin{figure}[htb]
\centering
\includegraphics[width=7cm]{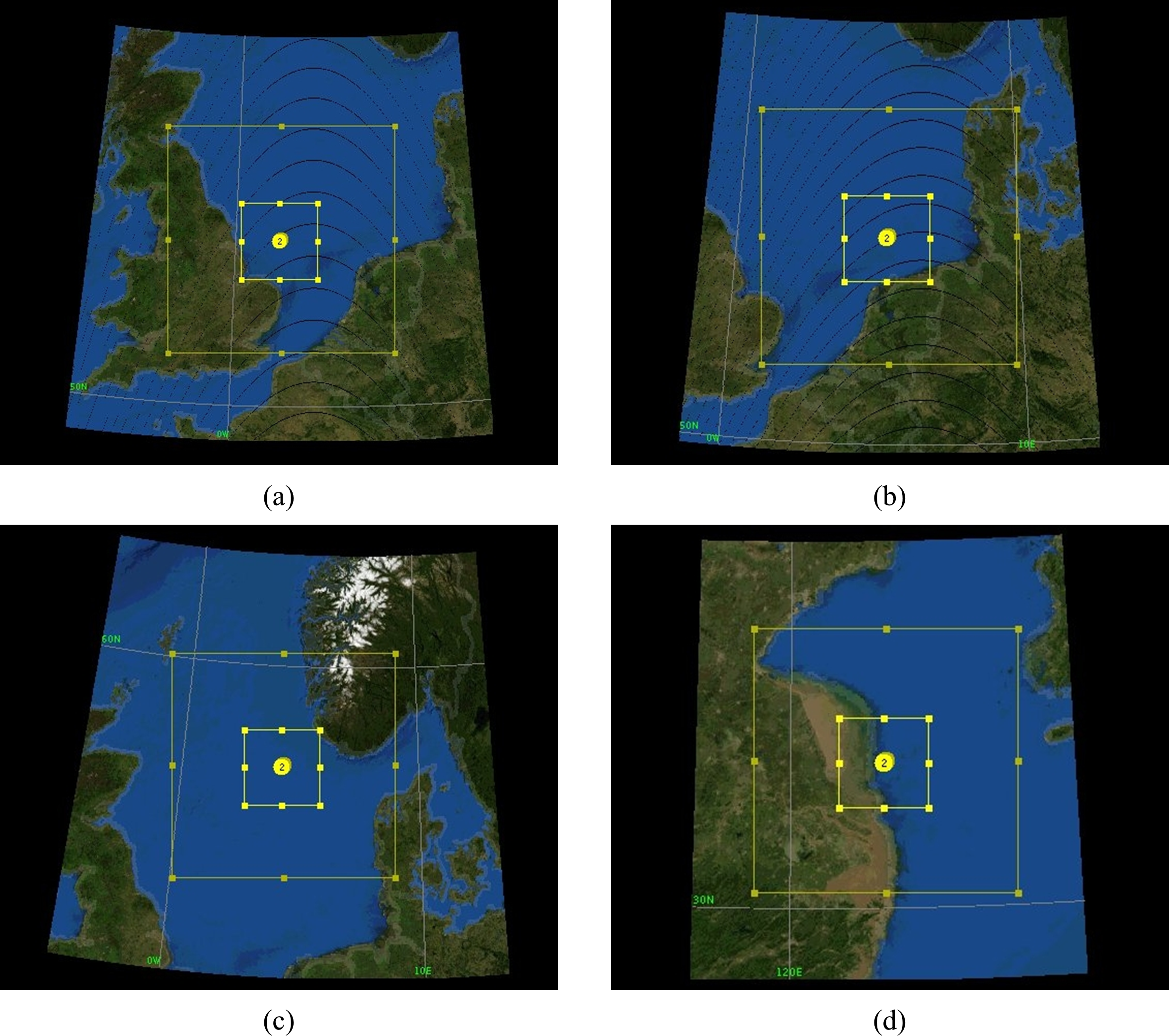}
\caption{Computational domains of the WRF simulation.}
\label{fig:1}
\end{figure}

For the wind speed without wind farms, the simulation based on WRF v4.3.1 is conducted. The simulation domains for the four selected wind farms are illustrated in Fig. \ref{fig:1}. Specifically, the outer domain encompasses 60 × 60 points with a horizontal grid spacing of 10 km × 10 km, while the inner domain consists of 101 × 101 points with a grid spacing of 2 km × 2 km. They are both configured with 40 vertical levels. The Global Final Analysis (FNL) data with $ 0.25^{\circ} $ resolution is employed as initial and boundary conditions to force the WRF model.

\subsection{Upstream and downstream areas}
\label{sec:uw}

Following the acquisition of observed and simulated wind speeds, the upstream wake-free and downstream wake-affected areas are manually labelled using the data based on the SAR-retrieved wind fields for subsequent comparison. Subsequently, upstream and downstream wind speeds are extracted from the SAR-retrieved wind field using the labelled data, with distances calculated based on the coordinates of the wind farms and points of interest within the wake-free and wake-affected areas. For WRF data, the simulation at the nearest hour to SAR is chosen as the reference of the `farm-less' scenario. The nearest grids in the WRF-simulated inner domain to the labelled wake-free and wake-affected pixels in SAR images are then identified to calculate the corresponding upstream and downstream wind speeds. After acquiring the distance to wind farms and the associated wind speeds, data averaging is performed based on one-kilometre units. 

\section{Results}
\label{sec:results}

\begin{figure}[htb]
\centering
\includegraphics[width=8.5cm]{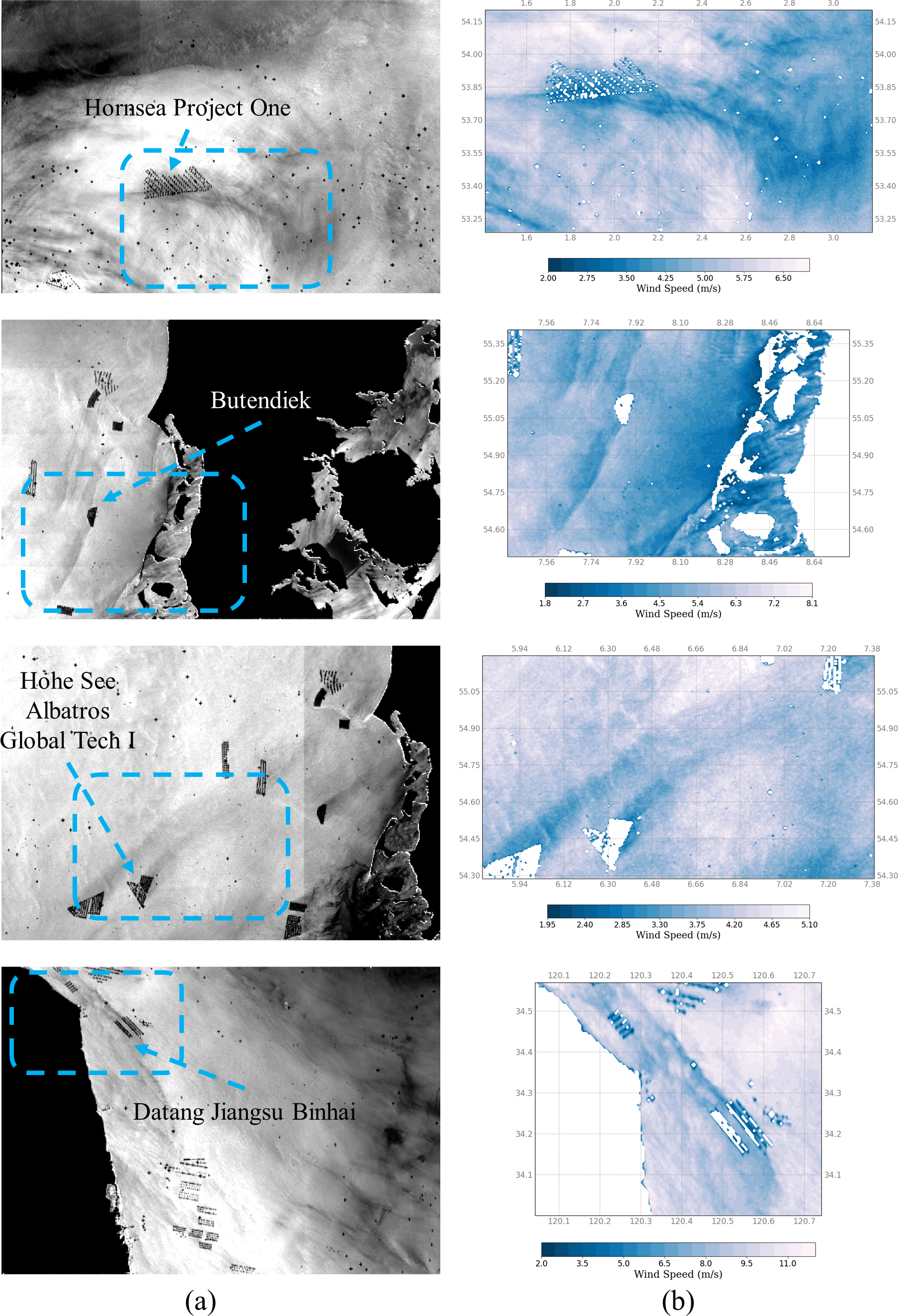}
\caption{The SAR-retrieved wind speed. The wind fields near wind farms of interest are enlarged in subfigure (b).}
\label{fig:2}
\end{figure}

\begin{figure*}[htb]
\centering
\includegraphics[width=17cm]{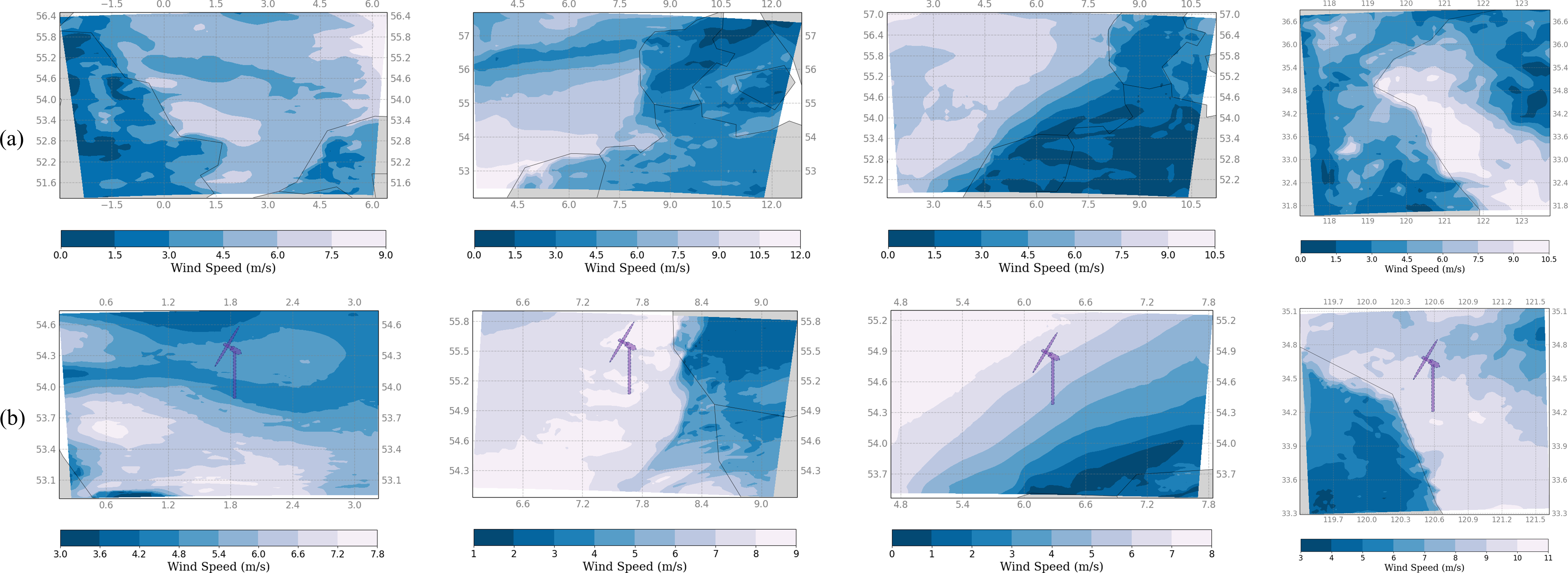}
\caption{The (a) outer domain and (b) inner domain of WRF-simulated wind speed, where purple turbines mark the locations of wind farms in the real world. Please note that the turbine marks are only used to indicate the farm locations to facilitate comparison with Fig \ref{fig:2}. The wind farm parameterization is not included in the WRF simulations.}
\label{fig:3}
\end{figure*}

\begin{figure*}[htb]
\centering
\includegraphics[width=15cm]{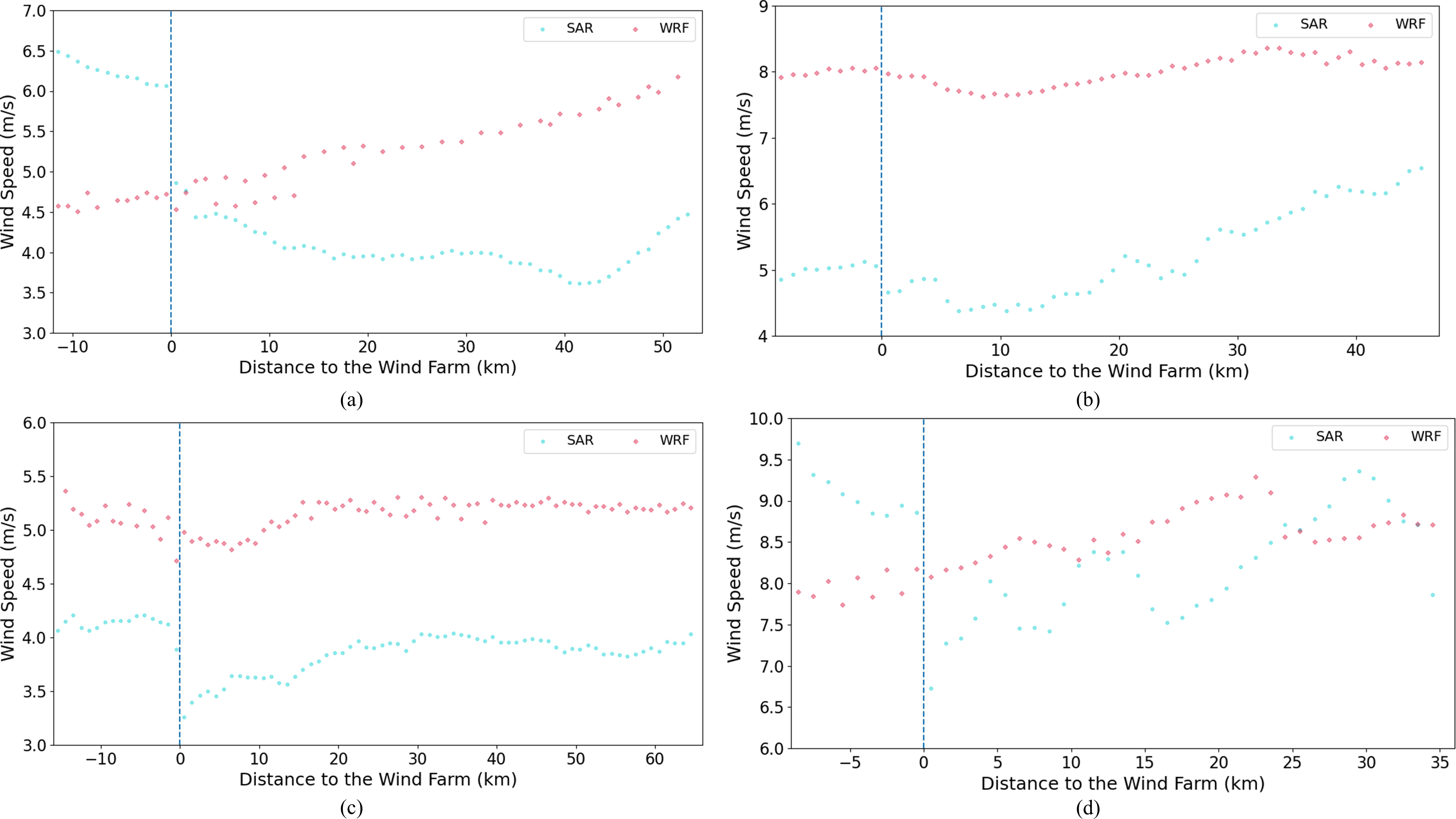}
\caption{Upstream and downstream wind speeds for four wind farm cases, including the SAR results for the scenarios with the presence of wind farms as well as the WRF results without wind farms.}
\label{fig:4}
\end{figure*}

The retrieved results from SAR are presented in Fig. \ref{fig:2}, with an enlarged view of the wind fields near the wind farms of interest shown in the subfigure (b). Notably, the wind farm wake effects are clearly visible in the enlarged figures, showcasing the wind speed reduction caused by the upstream wind farms. In Fig. \ref{fig:3}, we illustrate the WRF simulation results, encompassing both the outer and inner domains. As the time steps of SAR and WRF are not fully aligned and the output of CMOD5.N is the equivalent neutral wind, a direct comparison between SAR-retrieved and WRF-simulated wind fields is impractical. Instead, we opt for depicting the variation trend of wind speed from upstream to downstream, as illustrated in Fig. \ref{fig:4}. As we focus on qualitatively comparing the changing trend of wind speed, both WRF-simulated and SAR-retrieved wind speeds are simply kept at 10m altitude instead of interpolating them into the turbine operating altitude, i.e. usually about 100m.

As shown in Fig. \ref{fig:4}, while the four wind farm cases demonstrate distinct patterns, the presence of reduced wind speeds (i.e., wake effects) after the wind farms is clearly visible for all cases, particularly when compared with the WRF-simulated results. Taking the first case as an example, a substantial speed reduction is observed in the downstream area, contrasting with the WRF-simulated results, which exhibit an increased tendency in the same region. In the second and third cases, though the WRF-generated results also indicate reduced wind speeds, a notably larger reduction range is discernible in the SAR-retrieved results. In the fourth case, a transient peak emerges in the downstream range of 10 km to 15 km, potentially attributed to the unique turbine arrangement. As shown in Fig. \ref{fig:2}, the fourth wind farm is divided into two parallel parts, creating a gap through which the wind can easily pass. This is likely the contributing factor to this distinctive transient peak. These comparative observations between SAR and WRF results provide valuable insights into the intricate wake effects at various wind farm configurations.



\section{Conclusion}
\label{sec:conclusion}

In this paper, we combined the WRF model and SAR to visually demonstrate the wake effects caused by large offshore wind farms. The wind fields retrieved by SAR images were compared with the `farm-less' reference simulated by WRF. Through case studies encompassing four wind farms in diverse locations, a distinct pattern of wind speed reduction along the downstream area emerged, attributed to the presence of wind farms. The integration of SAR with WRF, as demonstrated in this study, underscores the considerable commercial potential of SAR in the realm of offshore wind applications.

It is noteworthy that the WRF-simulated results can only serve as a qualitative reference as it is not the real ground truth of the wind flows in the scenarios without the wind farm's impact, which is inherently non-existent. In our future work, we aim to further improve the understanding and quantification of inter-farm wake effects via integrating WRF with wind farm parameterizations into the technologies presented in the current paper.

\section{Acknowledgments}
\label{sec:acknow}
This work has received funding from the UK Engineering and Physical Sciences Research Council (grant number: EP/Y016297/1). The authors acknowledge the SCRTP at the University of Warwick for HPC resources. We express great appreciation to Copernicus Data Space Ecosystem for Sentinel images, Copernicus Climate Change Service for EAR5 data, European Space Agency for SNAP toolbox and National Centers for Environmental Prediction for FNL data.

\bibliographystyle{IEEEbib}
\bibliography{strings,refs}

\end{document}